\documentclass[11pt]{article}

\usepackage[left=2cm, right=2cm, top=2.5cm, bottom=2.5cm]{geometry}
\usepackage[x11names]{xcolor}
\usepackage{fancyhdr, amssymb, cancel, amsmath, graphicx, pgfplots, tikz}

\newcommand{\stylecolor}{black}

\usepackage[colorlinks=false, urlcolor=blue, linkcolor=blue, citecolor=violet, hyperindex=true, linktocpage=true]{hyperref}

\pgfplotsset{samples=200}

\usepackage[explicit]{titlesec}

\newcommand*\sectionlabel{}
\titleformat{\section}
  {\gdef\sectionlabel{}
   \Large\bfseries\scshape}
  {\gdef\sectionlabel{\thesection. }}{0pt}
  {\begin{tikzpicture}[remember picture,overlay]
	\draw (-0.2, 0) node[right] {\textsf{\sectionlabel#1}};
	\draw[thick] (0, -0.4) -- (\textwidth, -0.4);
       \end{tikzpicture}
  }
\titlespacing*{\section}{0pt}{15pt}{20pt}

\newcommand*\subsectionlabel{}
\titleformat{\subsection}
  {\gdef\subsectionlabel{}
   \large\bfseries\scshape}
  {\gdef\subsectionlabel{\thesubsection.\ \  }}{0pt}
  {\begin{tikzpicture}[remember picture,overlay]
    	\draw (-0.15, 0) node[right] {\textsf{\subsectionlabel#1}};
       \end{tikzpicture}
  }
\titlespacing*{\subsection}{0pt}{10pt}{10pt}

\pgfplotsset{every axis legend/.append style={at={(1.02,1)},anchor=north west}}

\newcommand{\titletext}{Multistable binary decision making on networks}

\begin{document}
\thispagestyle{empty}

\begin{equation*}
\begin{tikzpicture}
\draw (0.5\textwidth, -3) node[text width = \textwidth] {{\huge \begin{center} \color{\stylecolor} \textsf{\textbf{\titletext}} \end{center}}}; 
\end{tikzpicture}
\end{equation*}
\begin{equation*}
\begin{tikzpicture}
\draw (0.5\textwidth, 0.1) node[text width=\textwidth] {\large \color{black} $\text{\textsf{Andrew Lucas}}^{ab}\text{\textsf{ and Ching Hua Lee}}^b$};
\draw (0.5\textwidth, -0.5) node[text width=\textwidth] {\small $\;^a$ \textsf{Jefferson Physical Laboratory, Harvard University}};
\draw (0.5\textwidth, -1) node[text width=\textwidth] {\small $\;^b$ \textsf{Department of Physics, Stanford University}};
\end{tikzpicture}
\end{equation*}
\begin{equation*}
\begin{tikzpicture}
\draw (0.5\textwidth, -6) node[below, text width=0.8\textwidth] {\small 
We propose a simple model for a binary decision making process on a graph, motivated by modeling social decision making with cooperative individuals.  The model is similar to a random field Ising model or fiber bundle model, but with key differences on heterogeneous networks.  For many types of disorder and interactions between the nodes, we predict discontinuous phase transitions with mean field theory which are largely independent of network structure.  We show how these phase transitions can also be understood by studying microscopic avalanches, and describe how network structure enhances fluctuations in the distribution of avalanches.    We suggest theoretically  the existence of a ``glassy" spectrum of equilibria associated with a typical phase, even on infinite graphs, so long as the first moment of the degree distribution is finite.  This  behavior implies that the model is robust against noise below a certain scale, and also that phase transitions can switch from discontinuous to continuous on networks with too few edges.  Numerical simulations suggest that our theory is accurate. };  
\end{tikzpicture}
\end{equation*}
\begin{equation*}
\begin{tikzpicture}
\draw (0, -13.1) node[right] {\textsf{Correspond with: }\texttt{lucas@fas.harvard.edu}};
\draw (\textwidth, -13.1) node[left] {\textsf{\today}};
\end{tikzpicture}
\end{equation*}

\tableofcontents

\pagestyle{fancy}
\fancyhead{}

\fancyhead[L] {\textsf{\titletext}}
\fancyhead[R] {\textsf{\thepage}}
\fancyfoot{}

\section{Introduction}

Over the past decade there has been an explosion of interest in the statistical physics community into the behavior of simple statistical models on networks.  These models often lead to insights about qualitative behavior in social systems, as random graphs are a low order approximation to realistic social networks \cite{barratbook, castellano2}.    Decision making processes have long been studied as a simple example of such an application.  The voter model \cite{sood1}, along with variations with nonlinearities \cite{castellano} or other complications \cite{volovik, castello}, is a famous example, although only a model of consensus building.  The Axelrod model is an alternative model which exhibits equilibria with diversity of opinions \cite{axelrod, castellano3, vazquez, gonzalezavella}.   Other spin models or agent-based models have been proposed to study financial markets \cite{bornholdt, samanidou}.

Many of the above models do not predict a key phenomenon:  the presence of shocks, catastrophes and discontinuous phase transitions as external parameters are slowly tuned.    Often times, entirely new models have been proposed to account for this phenomenon \cite{watts, morenofm, crucitti}.   More interesting, however, is the proposal that the random field Ising model, well-known for hysteresis and discontinuous phase transitions \cite{sethna}, can be used to model these phenomena in social science \cite{anderson, jpb, galam}.   A similar model called the fiber bundle model, used to study the breakdown of some materials, also has similarly promising features \cite{pradhan, kimkim}.  Similarities between the fiber bundle model at a phase transition and the behavior of financial markets have also been noted \cite{voit}.   Other recent models have attempted to discuss disorder-induced phase transitions of opinion dynamics, using disorder in the interactions between individuals \cite{biswas}.   

In this paper, we propose a very simple model for a binary decision making process on a network.   Our model is similar to the random field Ising model in a global magnetic field, but with some important differences which make our model nearly exactly solvable on heterogeneous graphs.   A preliminary mean field analysis of the model predicts disorder-induced discontinuous phase transitions and hysteresis.   We then provide a microscopic justification for mean field theory as well, describing the microscopic dynamics of binary decision making in terms of avalanches.   We also describe how fluctuations in the sizes of avalanches can scale with the size of the network, on certain heterogeneous networks with fat tails.

Most interestingly, we will show that there is an infinite spectrum of equilibria in the large graph limit.  This is not surprising, because the random field Ising model has spin glass-like characteristics, often enhanced on networks \cite{sg1, sg2, sg3}.  Since our model does not admit a Hamiltonian and free energy,  the ``glassy" behavior of the binary decision model will be characterized by the presence of this spectrum of equilibria.  We will then use this spectrum of equilibria to justify two phenomena:  the robustness of the binary decision model to small fluctuations, and the possibility that network structure can suppress a discontinuity in the phase transition.   

The outline of our paper is as follows.  In Section \ref{sec2} we describe the binary decision model and provide some intuitive justification.   In Section \ref{sec3}, we describe a mean field analysis of the model, beginning with an exactly solvable case which has discontinuous phase transitions.   We then discuss numerical simulations, confirm that the model is roughly independent of the network structure, and discuss fluctuations in equilibria due to small network sizes.    Section \ref{sec4} describes avalanche dynamics, and provides a microscopic explanation for the independence of the mean field theory on network structure.   In Section \ref{sec5}, we describe ``multistability", which is the appearance of an exponential number of equilibria.   We propose this phenomenon first through a heuristic argument, and then through a more rigorous cavity calculation.   We conclude the paper by a discussion of basic consequences of multistability.

\section{The Binary Decision Model}\label{sec2}
In this section, we will introduce the binary decision model, justifying its use as a simple model for equilibrium social behavior.     We begin by approximating that individuals interact via a social network, which can be described as an undirected graph $G$ consisting of a vertex set $V$ and edges $E$.   We will denote $N=|V|$:  i.e., there are $N$ nodes in the graph.  We will denote the number of edges of a given node $v$ in the graph with $k_v$;   in mean field theory, we will often group together all nodes with the same number of edges, as is often done\cite{barratbook}.   To each node $v$ in the graph, we associate a binary variable $x_v \in \lbrace 0, 1\rbrace$.       For example, $x_v=0$ may mean that the individual is uninterested in participating and trading in a given economic sector, while $x_v=1$ means the opposite;  alternatively $x_v=0$ could model that an individual does not have an active account for a social media service, with $x_v=1$ the opposite.    Each node will decide its state, 0 or 1, by comparing an internal field, which we label $s_v$, to an external (global) field $p$:\footnote{It is not important what happens when $s_v=p$ for almost all reasonable formulations of the graph $G$, or the internal fields $s_v$.} \begin{equation}
x_v = \Theta(s_v-p).   \label{psv}
\end{equation}
For example, if we think of $p$ as the external price of some good, then $s_v$ represents the effective price at which buyer $v$ is willing to buy:  when $s_v>p$, $x_v=1$ and the buyer is actively buying, and when $s_v<p$, $x_v=0$ and the buyer is not actively buying.

In order to fully specify the model we thus simply need to describe how to determine $s_v$.    Our formulation of the binary decision model will be to assume that \begin{equation}
s_v \equiv P_v h(q_v) \label{spv}
\end{equation}where $P_v$ is some internal variable, $h$ is a monotonically increasing function, and\begin{equation}
q_v = \mathrm{P}(x_u = 1|uv\in E)
\end{equation}simply represents the fraction of neighbors of $v$ which are in state 1.  We are free to rescale $P_v$ so that $h(0)=1$, and we will assume so for the remainder of the paper.     Note that we will always assume that $h$, $P_v$ and $p$ are non-negative.   Note that under the identifications \begin{subequations}\begin{align}
p &= \mathrm{e}^{p_+}, \\
P_v &= \mathrm{e}^{P_{+v}}, \\
h(q) &= \mathrm{e}^{h_+(q)},
\end{align}\end{subequations}we can formally write the binary decision model as \begin{equation}
x_v = \Theta(h_+(q_v) + P_{+v} -p_+),
\end{equation}since the exponential is a monotonically increasing function.

To understand the choice above, it is helpful to re-cast the problem temporarily in the language of economics.   Suppose that $p$ represents some globally observed price (or, more abstractly, ``utility" parameter);  then, $P_v$ represents the price/utility that node $v$ believes is the value of existing in state 1.   The key point of this model is that the effective value of $P_v$ is multiplied when neighbors of $v$ are also in state 1.   For example, a social media service (where $P$ represents a utility, not a specific price) is far more valuable to its users when there are many other users.    Similarly, stock traders may base much of their valuation of a stock based on what they believe the rest of the market is doing.    Therefore, this binary decision model represents social scenarios where individuals are making a binary decision based on comparing the utility of two options, when the utility of one option is dependent on the behavior of their neighbors.   Furthermore, the decision making is \emph{cooperative} in the sense that if node $v$ switches from $x_v=0$ to $x_v=1$, the probability that any neighbor of $v$ is in state $x=1$ is non-decreasing.   This is contrast to \emph{antagonistic} decision making, where the probability that any neighbor of $v$ is in state $x=1$ is non-increasing after $v$ switches state.   In this paper, we will only consider the cooperative case, where the individuals tend to make the same decisions as their neighbors.

The argument can and has been made \cite{jpb} that an alternative choice, \begin{equation}
s_v = P_v + J\sum_{(uv)\in E} x_u,  \label{svrfim}
\end{equation}which directly results in the famous random field Ising model, is also worth studying.  In fact, note that this model is formally equivalent to the binary decision model on graphs where all nodes have the same number of edges, for a special choice of $h_+(q) = kJq$.   For the purposes of this paper, we will stick with (\ref{spv}), which leads to simpler calculations.    We should note, however, that there is one major drawback to a choice such as this:  as far as we can tell, there is, in general, no Hamiltonian function which will result in the binary decision model, and this denies us the use of some of the tools of statistical mechanics.

\section{Mean Field Theory}\label{sec3}
Let us now turn to mean field theory to ``solve" the binary decision model, and compare to simulations.   Our first pass will demonstrate both how mean field theory can be quite accurate, as well as reveal some of its major drawbacks.
\subsection{A Macroscopic Solution}
Let us denote \begin{equation}
q = \mathrm{P}(x_v = 1)
\end{equation}where the average over nodes is over the uniform distribution over $V$.   Let us also assume that the random variables $P_v$ are i.i.d. drawn from a probability distribution with cumulative distribution function $1-F(P)$, and probability density function $f(P)$.  It is straightforward to see that \begin{equation}
q = \mathrm{P}(h(q) P_v > p) = F\left(\frac{p}{h(q)}\right) \label{qeq}
\end{equation}where the mean field approximation that $q_v=q$ has been applied.  (\ref{qeq}) is the mean field equation describing the possible equilibria of the system, as described by the single parameter $q$.   Note that this mean field equation appears independent of the degree distribution of the graph, if we chose to take this into account, since \begin{equation}
q_k = \mathrm{P}(x_v = 1|k_v=k)=\mathrm{P}(x_v = 1)=q.
\end{equation}The middle equality above is a consequence of the fact that in mean field theory, every edge points to the same ``mean field node":  i.e., every edge contributes the same factor of $q$.    There are cases where this approximation will break down, and we will return to this at the end of the paper.

\begin{figure}[here]
\centering

   \includegraphics{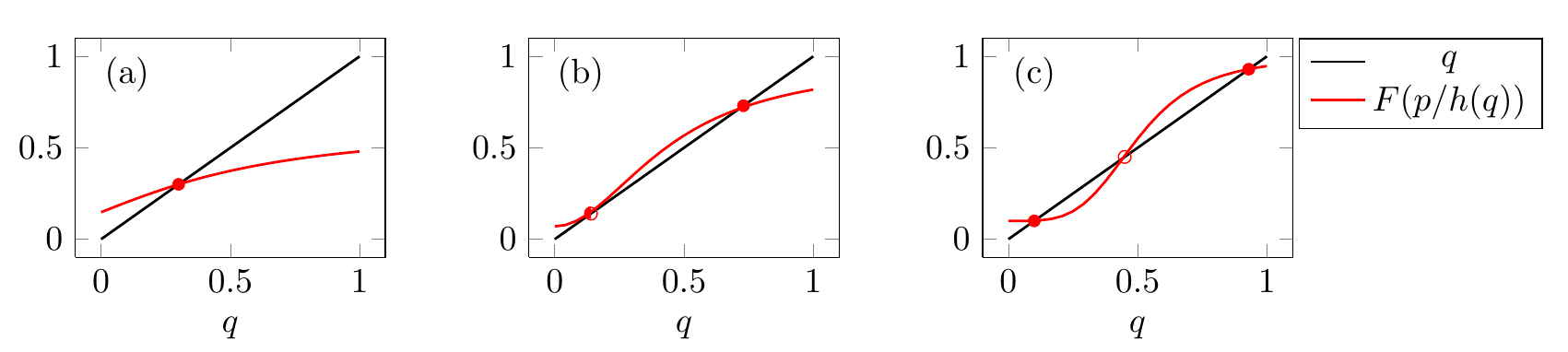} 
   
\caption{The graphical method of solving (\ref{qeq}), in various cases.   (a) shows a case with one solution;  (b) shows a critical point;  (c) shows a bistable case with three solutions.   The solid dots represent equilibria which are stable , and the empty dots represent equilibria which are unstable:  the half-filled dot represents a marginal point.}
\label{graphexplain}
\end{figure}

We are most interested in the case where there are multiple equilibria, i.e. where (\ref{qeq}) has multiple solutions.   Graphically it is clear how to approach this problem, as shown in Figure \ref{graphexplain}.  Inspired by this approach, we can also straightforwardly analyze it analytically.  Since $F$ is a CDF, we have $0\le F \le 1$.  Because we must have $0 \le F(q=0)$ and $1 \ge F(q=1)$, there must be some $q^*$ for which (\ref{qeq}) is true.  Furthermore, let us consider the behavior of $F$ near this point.   Suppose that $\mathrm{d}F/\mathrm{d}q >1$ -- this means that $F<q$ for $q$ just smaller than $q^*$, and $F>q$ for $q$ just larger than $q^*$.   However, given the constraints at $q=0$ and $q=1$, we see that there must be at least 2 other points at which $q=F$, implying the existence of at least 3 solutions to (\ref{qeq}).   We conclude that if \begin{equation}
\left.\frac{\mathrm{d}F(p/h(q))}{\mathrm{d}q}\right|_{q=q^*} = f\left(\frac{p}{h(q)}\right)\frac{ph^\prime(q)}{h(q)^2} \equiv \alpha  > 1,
\end{equation}then we have multiple equilibria.   In a later section, we will find a simple microscopic interpretation for $\alpha$ as the probability that a spontaneous flip in any node's state will cause one of its neighbors to also flip, and the stability criterion $\alpha<1$ follows from the condition that an avalanche have finite expected size.

Now, let us consider what happens as we change $p$.  In particular, suppose we increase $p$ to some $p=p_{\mathrm{c}}$ for which \begin{equation}
f\left(\frac{p_{\mathrm{c}}}{h(q^*)}\right) \frac{p_{\mathrm{c}}h^\prime(q^*)}{h(q^*)^2} = 1.   \label{eq3}
\end{equation}Then we conclude that passing through $p_{\mathrm{c}}$ in the direction which decreases the left hand side of (\ref{eq3}) corresponds to the disappearance of a pair of fixed points.   In particular, if the solution at $q^*$ disappears, it must discontinuously jump to another point.   This is the hallmark of a discontinuous phase transition.   It is not always the case that a distribution of $F(P)$, and an interaction $h(q)$, will allow such a discontinuous phase transition, but quite often discontinuous phase transitions do occur.   Essentially, increasing the value of $h(q)$ makes the model more and more likely to admit a discontinuous phase transition, and subsequently increase the size of the jump at the transition.
\subsection{An Exactly Solvable Case}
Let us look at a sample of an exactly solvable version of this model, to the level of approximation we have just studied.   We will take \begin{equation}
h(q) = 1+Aq.  \label{linearh}
\end{equation}and \begin{equation}
F(P) = \left\lbrace \begin{array}{ll} 1 &\ P<P_0 \\ P_0+1-P &\ P_0<P<P_0+1 \\ 0 &\ P>P_0+1 \end{array}\right..
\end{equation}

It is easy to check when $q=0$ is a solution:  this occurs when $F(p/h(0)) = 0$, or $F(p)=0$, or $p\ge P_0+1$.    $q=1$ is a solution if $F(p/(1+A)) = 1$, or $p<(1+A)P_0$.   For $0<q<1$, solutions occur when \begin{equation}
q = 1+P_0 - \frac{p}{1+Aq},
\end{equation}which can easily be solved using the quadratic formula to give \begin{equation}
q = \frac{1}{2}\left[P_0 + 1-\frac{1}{A} \pm \sqrt{\left(P_0+1+\frac{1}{A}\right)^2 - \frac{4p}{A}}\right]  \label{eq9}
\end{equation} 
The important feature of (\ref{eq9}) is the square root, which will become imaginary at price \begin{equation}
p_{\mathrm{c}} = \frac{A}{4}\left(P_0+1+\frac{1}{A}\right)^2.
\end{equation}

\begin{figure}[here]
\centering

   \includegraphics{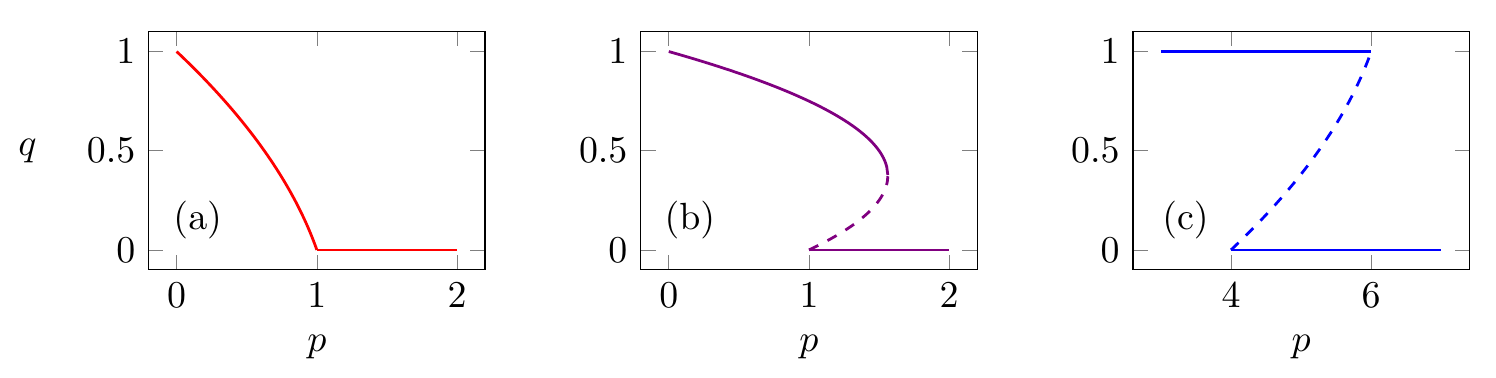} 
   
\caption{Three examples of exact mean field solutions for uniform distributions on $P_v$ displaying qualitatively different behaviors.   (a) corresponds to $P_0=0$, $A_0=0.5$;  (b) corresponds to $P_0=0$, $A=4$;  (c) corresponds to $P_0=3$, $A=1$.    The solid lines represent stable equilibria, and the dashed lines unstable equilbria -- we will understand stability from a microscopic standpoint later.}
\label{3exs}
\end{figure}

There are 3 distinct qualitative possibilities for the phase diagram.    To understand these possibilities, it will suffice to consider $q$ at $p=p_{\mathrm{c}}$, which is \begin{equation}
q_{\mathrm{c}} = \frac{1}{2}\left( P_0 + 1 -\frac{1}{A}\right).
\end{equation}
If $q_{\mathrm{c}}>1$,  which occurs when $P_0 > 1+A^{-1}$, then we know that only one branch of physical solutions exists (other than $q=0$ or 1), and this branch has a positive slope in the $pq$ plane, connecting $p=(1+A)P_0$ and $p=P_0+1$.    If $q_{\mathrm{c}}<0$, then one branch of physical solutions exists, with negative slope, connecting the same 2 points.   Otherwise, we see that $q_{\mathrm{c}}$ is a physical value, and therefore is a valid point in parameter space -- in particular, there are two branches of allowed solutions (other than $q=0$ or 1).   We show examples in Figure \ref{3exs}.

Physically, we see that depending on the  parameters, the binary decision model has a variety of interesting behaviors.   When $A$ is small enough, there are no discontinuous phase transitions, suggesting ``nice behavior".   However, when $A$ gets large enough, discontinuous transitions begin to occur, and when $A$ reaches a critical value, the only stable equilibria are with the entire graph in state 0 or 1, representing a hyper-polarized situation.
\subsection{Numerical Simulations}
We now present numerical simulations of the binary decision model.   In our numerical simulations, we always take $h(q)$ to have the linear form  (\ref{linearh}), although we allow the coefficient $A$ to vary.   We ran simulations over  Erd\"os-R\'enyi random graphs, drawn from an ensemble where every possible edge is equally likely to be included in the graph.   We then repeated simulations on asymptotically scale free graphs, generated by a modification of the algorithm of Ref. \cite{krapivsky} in which the added nodes have multiple edges.  The algorithm is particularly easy to implement numerically, generates graphs which have degree distributions with robust scale free ($\rho_k\sim k^{-\nu}$) tails,   and can generate distributions for an arbitrary exponent $\nu>2$.    It does have the peculiarity that if each node that is added has $m$ edges (in the large $N$ limit), $\langle k\rangle = 2m$, although this is just signifying that the small $k$ distribution is not scale free.   Since we are interested in studying the model on scale free graphs for any possible effects of nodes with very large $k$, this peculiarity is acceptable.   We also note that many of our theoretical results are robust against the details of the graph ensemble, and our simulations confirm this claim.

    For our internal $P$ distributions, we generated $P_v$ from either uniform distributions over $[0,1]$, or Gaussian distributions with $\mu=1$, $\sigma$ varying around $0.2$, and scale free distributions with varying exponents and minimum $P_v$ of 1.   We chose these distributions because they represent three different types of behavior, which could cause our mean field theory approximations to break down.    The uniform distribution has substantial fractions of nodes with very low $P_v$; in contrast, the scale free distributions have nodes with very large $P_v$.   The Gaussian distribution tends to cluster nodes around $P_v\approx 1$.    To observe bistability if it exists, we began our simulation by starting with $p=0$, and then increased $p$ in uniform steps up until some value $p_{\mathrm{max}}$, and then decreased $p$ in equal steps back to $p=0$.

\begin{figure}[here]
\centering

   \includegraphics{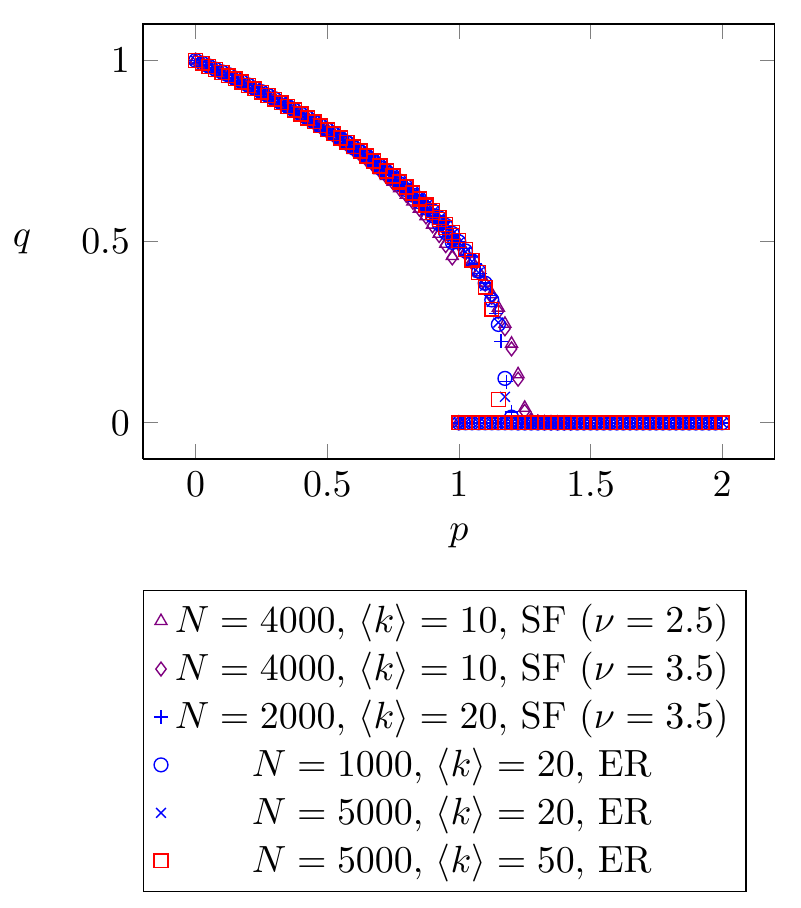} 

\caption{The average value of $q(p)$ for a uniform distribution of prices with $A=2$.  We averaged over at least 50 trials for each type of network.   The smoother transitions correspond to graphs with smaller $\langle k\rangle$, a fact which we will explain later: note that there is only a single square ($\langle k\rangle = 50$) in the crossover regime, but many triangles ($\langle k\rangle = 10$)}
\label{varynandm}
\end{figure}

\begin{figure}[bottom]
\centering

   \includegraphics{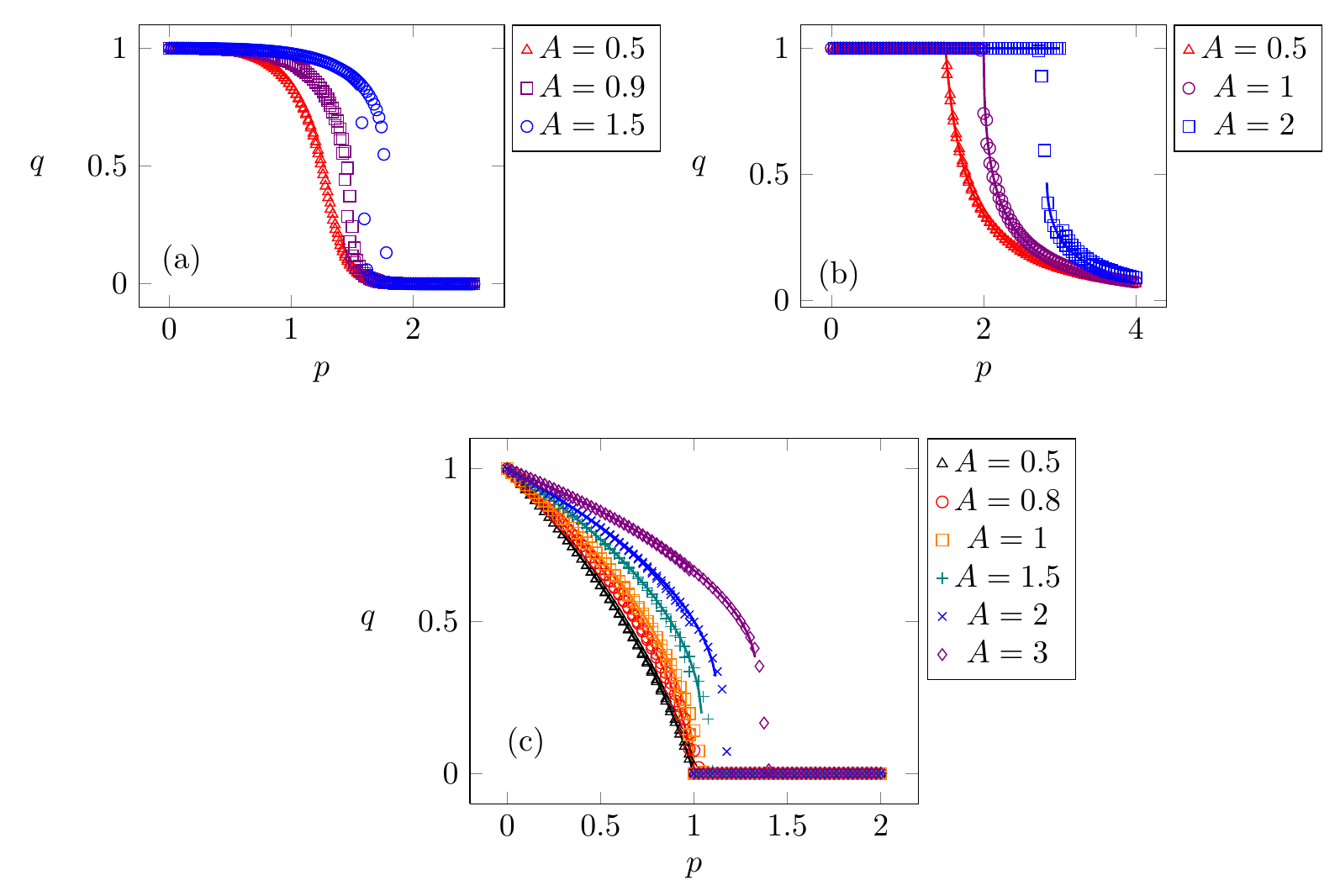} 

\caption{Here we show how increasing the ``interaction strength" in the binary decision model causes the onset of a discontinuous phase transition, both in theory and simulation.   Our simulations for this graph used Erd\"os-R\'enyi graphs with 5000 nodes and $\langle k\rangle = 20$.   (a):  the distribution of $P_v$ is Gaussian with mean 1 and standard deviation $\sigma=0.3$;  (b):  $P_v$ distribution is scale free with exponent $\nu=3$;  (c):  $P_v$ distribution is uniform on [0,1].   Note that the discontinuous phase transition appears smeared out because of fluctuations in the realizations of disorder;  data from individual runs clearly indicates the presence of discontinuous phase transitions for larger values of $A$.}
\label{intst}
\end{figure}

As a first simulation, we demonstrate that for a given internal disorder distribution $F(P)$, and a given $h(q)$, the mean field theory solution of Eq. (\ref{qeq}) is typically a very good approximation.   This is shown in Figure \ref{varynandm}.   Note that typically, the phase transition does not look as sharp as it occurred because we are averaging over runs, and there are small fluctuations in the critical value of $p$ due to internal disorder.   We emphasize that the details of the graph ensemble, beyond $\langle k\rangle$, appears to have no effect on the curve $q(p)$.    Figure \ref{intst} shows the emergence of a phase transition as the interaction strength $A$ is increased past a critical value of 1, for the given uniform price distribution, as shown by both theory and numerics, as well as some sample distributions where $P_v$ is a Gaussian or scale free random variable.   In each case, up to the deviations from mean field theory described above, we see excellent agreement with the theory, suggesting that for any $P_v$ distribution, mean field theory will be a good approximation.

While the mean field theory approximation appears to typically hold, we should point out some \emph{key} failures.  In particular, there is a noticeable hysteresis effect associated to the upper branch of the mean field solution near the phase transition.   We also notice that the phase transition appears to be slightly delayed for smaller values of $\langle k\rangle$, and on some graphs it appears as though there is no discontinuous phase transition at all.   Furthermore, this effect cannot be removed by increasing $N$.   Both of these phenomena are key signatures of the advertised multistability, and we will return to them later.

\subsection{Finite Size Fluctuations from Disorder}
Let us briefly discuss the fluctuations about mean field theory due to finite size effects (but \emph{not} small network effects).   These fluctuations are simply due to the fluctuations in the values of the internal fields, $P_v$.   To quantitatively estimate their size,  let us denote $q=q_0+\delta$, with $q_0$ the mean field value predicted by (\ref{qeq}) and $\delta$ a small fluctuation.   Similarly, let us denote the CDF of the actual distribution of $P$, realized on the graph, as $F=F_0+\Delta$, with $F_0$ the mean field value and $\Delta$ a small fluctuation.   Then, expanding (\ref{qeq}) to lowest order in the fluctuations we find \begin{equation}
q_0+\delta = F_0\left(\frac{p}{h(q_0+\delta)}\right) + \Delta\left(\frac{p}{h(q_0)}\right) = q_0 + \alpha \delta + \Delta\left(\frac{p}{h(q_0)}\right),
\end{equation}which implies that \begin{equation}
\delta = \frac{\Delta}{1-\alpha}.
\end{equation}

\begin{figure}[here]
\centering

   \includegraphics{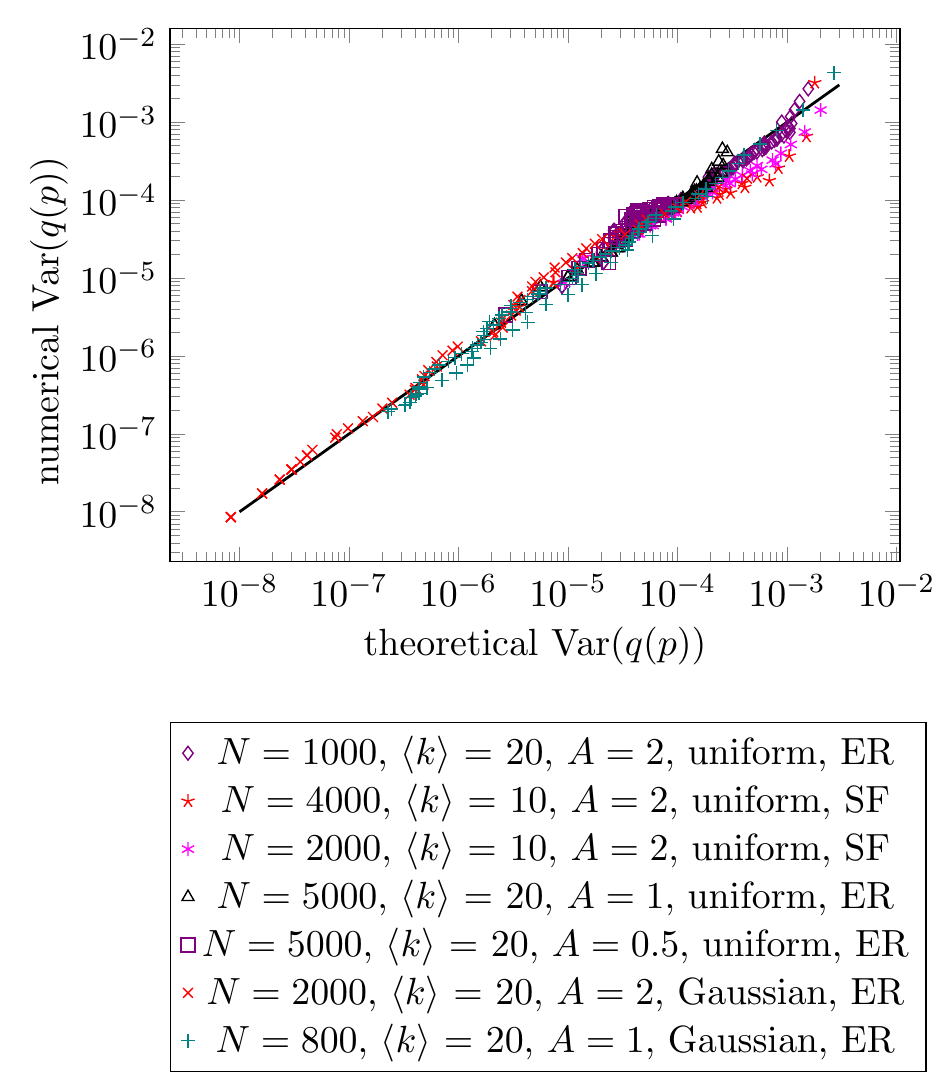} 

\caption{We compare the predictions of Eq. (\ref{varqeq}) to numerical simulations on a variety of graphs.   The last two entries of the legend refer to the distribution of $P_v$ and the graph type, respectively.   For the data shown, Gaussian distributions have $\sigma=0.3$ and scale free graphs have $\nu=2.5$.    For clarity, only some of our data is shown, although we emphasize that the data not shown was just as good of a fit.  Significant deviations come near the onset of a phase transition -- we have removed some of these values near the critical point when our code is averaging over realizations in different phases.}
\label{sec51f}
\end{figure}

The distribution of $\Delta$ is simple to find, although we will focus only on the variance of the fluctuations, as the higher order fluctuations are rapidly suppressed for increasing $N$.  Since the internal disorder consists of iid random variables, and for the given value of $p$ and the distribution of $P_v$ the probability that a node is in state $x_v=1$ is $F_0$, assuming the rest of the network to be in a mean field state, we conclude \begin{equation}
\mathrm{Var}\left(\frac{1}{N}\sum x_v\right) = \frac{1}{N^2}\sum \mathrm{Var}(x_v) = \frac{F_0(1-F_0)}{N} = \frac{q(1-q)}{N}.
\end{equation}This implies that \begin{equation}
\mathrm{Var}(q) = \frac{q(1-q)}{(1-\alpha)^2}\frac{1}{N}.  \label{varqeq}
\end{equation}
Since this number is typically quite small due to the factor of $1/N$, and our theory suggests that this variance is not dominated by large deviations, we postulate that higher order fluctuations will not play an important role.  We find that this relation is obeyed very well so long as we do not approach $\alpha=1$, as shown in Figure \ref{sec51f}.   

\section{Avalanches}\label{sec4}
We now turn to a slightly different question, which is a reinterpretation of our previous mean field analysis on the existence of a phase transition.   Let us consider a state of the graph in equilibrium at $q^*=F(p_0/h(q^*))$.   Suppose that the external field $p_0$ is increased slightly to $p$, causing only node $v_0$ to flip.   How many other spins will also flip, due to the change of state of $v_0$?   

Let us define $Q$ to be the probability that one of the vertices $v$ connected to $v_0$ will flip: \begin{align}
Q &= \mathrm{P}\left(\frac{p}{h(q_v - 1/k_v)} > P_v > \frac{p}{h(q_v)}\right) = F\left(\frac{p}{h(q)}\right) - \left\langle F\left(\frac{p}{h(q-1/k)}\right)\right\rangle_{\mathrm{edges}}  \notag \\
&= F\left(\frac{p}{h(q)}\right) - F\left(\frac{p}{h(q)}\right) + \left\langle \frac{1}{k} f\left(\frac{p}{h(q)}\right)\frac{ph^\prime(q)}{h(q)^2} \right\rangle_{\mathrm{edges}} = \alpha \left\langle \frac{1}{k}\right\rangle_{\mathrm{edges}}. 
\end{align}Now, we have to be a bit careful.   As noted above, the averaging occurs over the distribution of nodes which an edge points to, which is different from the distribution of nodes itself, because nodes with more edges are counted more often:  in fact, each node $v$ is counted $k_v$ times in the average $\langle \cdots \rangle_{\mathrm{edges}}$ over nodes pointed to by an edge.   Therefore:\begin{equation}
\left\langle \frac{1}{k}\right\rangle_{\mathrm{edges}} = \sum \frac{k\rho_k}{\langle k\rangle} \frac{1}{k} = \frac{1}{\langle k\rangle},
\end{equation}so we conclude that \begin{equation}
Q = \frac{\alpha}{\langle k\rangle}.  \label{eqq}
\end{equation}

Let us now denote with $n$ the number of the vertices $v$ we expect to transition to the 0 state.   For any given site, this occurs with probability $Q$, so \begin{equation}
\langle n\rangle = \left\langle \sum_{j=1}^{k_{v_0}} Q\right\rangle = \sum \rho_k kQ = \langle k\rangle Q = \alpha.
\end{equation}
Assuming that $Q$ stays constant, if any new spin flips, then it too will have the possibility of flipping spins.     If we approximate the new spin as the same as the old spin\footnote{Arguably, the $k$ above should be replaced with a $k-1$, but for $k\gg 1$ this is not a major qualitative or quantitative change.  We also require $k\gg 1$ for our Taylor approximation above to be accurate, so we will for simplicity neglect worrying about this for this paper.} then, since the avalanche approximately grows as a birth/death process, we conclude the total size of the avalanche is \begin{equation}
\langle n_{\mathrm{total}}\rangle \approx \frac{\alpha}{1-\alpha}.
\end{equation}Of course, this formula only holds for $\alpha<1$; for $\alpha> 1$, it is well known that the birth/death process has an expected infinite size, and for $\alpha=1$ it is almost surely finite but with infinite expected value.

This gives us much new insight into the physical processes at work behind mean field theory.  We see looking back at (\ref{eq3}) that we found that reaching a point where $\alpha=1$ corresponds to a phase transition -- here, that means that we expect a spin avalanche to have infinite size.    Furthermore, it allows us to analyze the stability of the fixed points we found earlier -- only the fixed points where $\alpha<1$ are stable.  Understanding phase transitions by considering microscopic avalanche sizes is not new:  see, e.g., \cite{jpb} in the context of the random field Ising model, or \cite{pradhan, kimkim, zapperi, kloster} in the context of the fiber bundle model.    Interestingly, however, for our binary decision model, this calculation helps to give insight into why the graph structure is seemingly so unimportant in mean field theory -- even though more connected nodes are affected more often by spin flips, they are not affected as much, and these effects cancel each other.   This would \emph{not} be the case in the random field Ising model on a heterogeneous network, for example, where $\alpha$ itself would obtain an intricate dependence on the degree distribution.   In this sense, our binary decision model, where $h(q)$ is $k$-independent, is a very convenient toy model to solve.

\begin{figure}[here]
\centering

   \includegraphics{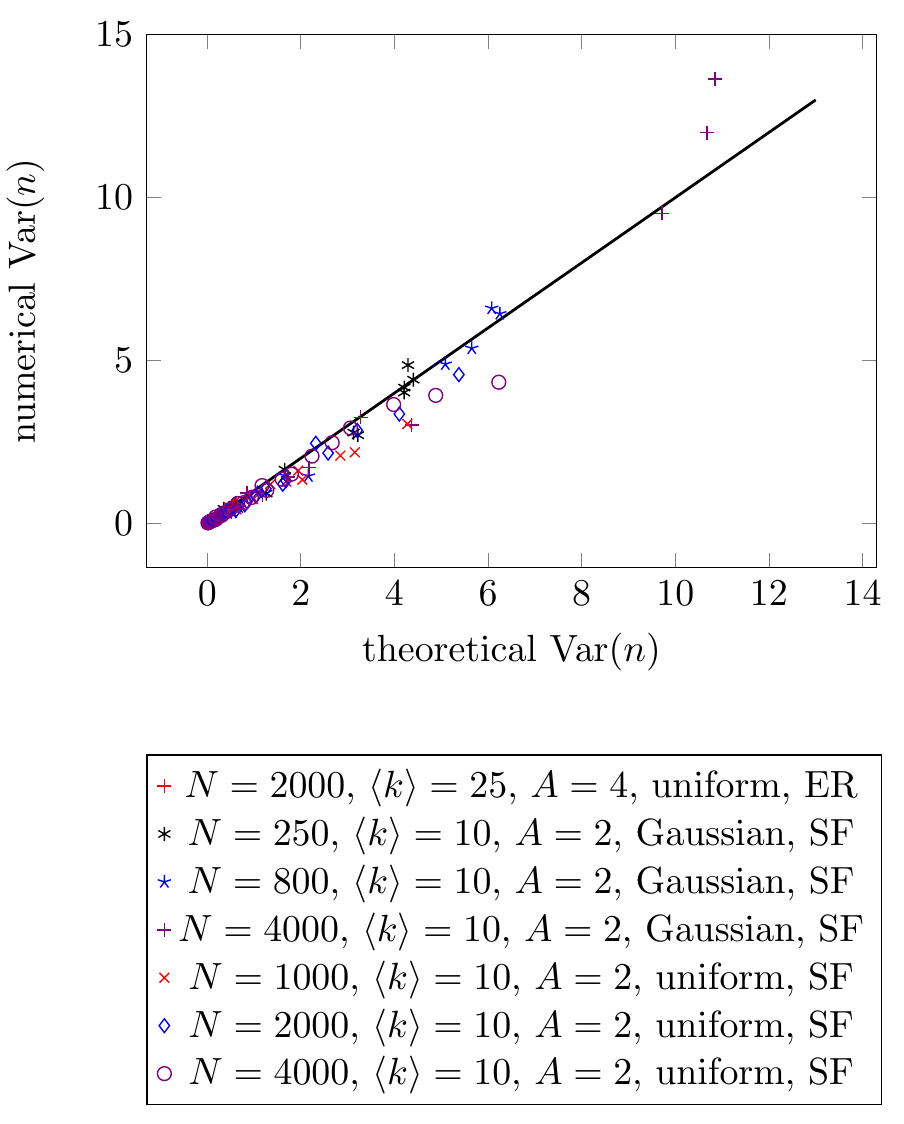} 

\caption{We compare theory vs. numerics for the predicted variance in the number of nodes which change state based on the change of a single node.   The deviations appear to become much more significant as we approach a phase transition, perhaps due to loopy effects in the graph (where our avalanche theory breaks down).   In the data shown, scale free graphs we used had $\nu=2.5$ and Gaussian $P_v$ distributions had $\sigma=0.2$.  We generated more data that appears very similar, but have not shown all of it for clarity.}
\label{perspinavagraph}
\end{figure}

Let us also ask about the fluctuations in the size of avalanches, which we will explore by considering variations in the size of $n$, the number of neighbors that one site flipping will also flip.    These fluctuations occur both within the same realization of graph structure and internal disorder.   However, for simplicity we have included many realizations in our average.   Denoting $z_v = 1$ if $v$ flips and 0 otherwise:  \begin{align}
\mathrm{Var}(n) &= \left\langle n^2 \right\rangle - \langle n\rangle^2 = \sum \rho_k \left\langle \left(\sum_{j=1}^k z_j\right)^2\right\rangle    -\alpha^2  = \sum \rho_k \left\langle\sum_{j=1}^k z_j^2 + \sum_{j\ne l} z_j z_l \right\rangle-\alpha^2 \notag \\
&= \sum \rho_k \left[k Q + k(k-1)Q^2\right] - \alpha^2 = \alpha(1-\alpha) + \frac{\langle k^2-k\rangle}{\langle k\rangle^2} \alpha^2.
\end{align}
We see that the graph structure thus plays an important role in fluctuations of the size of avalanches.  This theory is quite accurate as shown in Figure \ref{perspinavagraph}, although it does seem to break down very close to a phase transition, as the diverging curves suggest.

Other work \cite{zapperi, kloster} considers in more detail the theory behind avalanche distributions in the fiber bundle model, which is quite similar to the binary decision model, on fully connected networks, in particular near the critical point, although there is no general theory of the distribution of avalanches accounting for heterogeneous network structure.    Indeed, our results above, as we saw that the variance included network structure even assuming a ``mean field" network structure.

\section{Multistability}\label{sec5}
In this section, we will discuss the binary decision model's most interesting feature -- ``multistability."  By the term multistability, we mean that there is a continuous spectrum of equilibrium states, even in the $N\rightarrow\infty$ limit, so long as $\langle k\rangle$ is finite.   Recall that an equilibrium for the binary decision model is a state where all nodes satisfy the constraint equation that $x_v = \Theta(P_vh(q_v)-p)$.   Since $q_v$ depends on the values of $x_u$ for each neighbor $u$ of node $v$, it is possible that there are multiple possible solutions to the constraint equation -- thus, multiple equilibria exist.   We've trivially seen that this can be true by the existence of two phases, but here we are interested in the possibility that multiple equilibria may exist for any given phase.

Due to the internal disorder, and the similarity to the random field Ising model, it is not surprising that glass-like behavior arises, although the lack of a formal Hamiltonian or free energy makes it challenging to classify the binary decision model as a glass.  Instead, we will call the behavior multistability.   The goal of this section is to propose  multistability from a theoretical standpoint.  We will then show it exists in our numerical simulations, and comment on the implications of this phenomenon.
\subsection{A Spectrum of Equilibria}
 Let us approximate the probability that a pair of spins will satisfy the $x_v$ constraint either if they are both in state 1 or in state 0.        Note that it will never be the case that we would need to consider the possibility that a pair of nodes could flip between 10 and 01, because the constraint equation implies that decisions are always made cooperatively.\footnote{With more work, this can be made rigorous.}  We will assume that the remainder of the graph is treated within the mean field approximation.  The probability that one node flipping will flip another node has been calculated in the previous section in (\ref{eqq}).   Thus, we simply calculate the probability that two nodes can be in either state 1 or 0 to be the probability node $u$ flipping flips node $v$ and vice versa, which is simply \begin{equation}
\mathrm{P}(\text{pair can exist in 2 states}) = \left(\frac{\alpha}{\langle k\rangle}\right)^2.
\end{equation}Since there are $N\langle k\rangle /2$ edges in the graph, we thus approximate the number of equilibria as being \begin{equation}
N_{\mathrm{equilibria}} \sim 2^{N\alpha^2/2\langle k\rangle}.
\end{equation}However, we now need to take into account the expected value of the avalanche caused by the pair flip.   This can be accounted for by roughly multiplying by a factor of $(1-\alpha)^{-1}$, corresponding to the expected size of the avalanche caused by one of the nodes.  Thus, we find that given a pair of nodes, we should expect that flipping them will cause a change of\begin{equation}
\Delta q_{\text{one pair}} \sim \frac{1}{N} \frac{2\alpha^2}{(1-\alpha)\langle k\rangle^2},
\end{equation}and this leads to a width of the equilibrium spectrum which is finite even in the $N\rightarrow\infty$ limit, so long as $\langle k\rangle$ is finite: \begin{equation}
\Delta q \sim \frac{\alpha^2}{(1-\alpha)\langle k\rangle}.  \label{1overk}
\end{equation}

\begin{figure}[here]
\centering

   \includegraphics{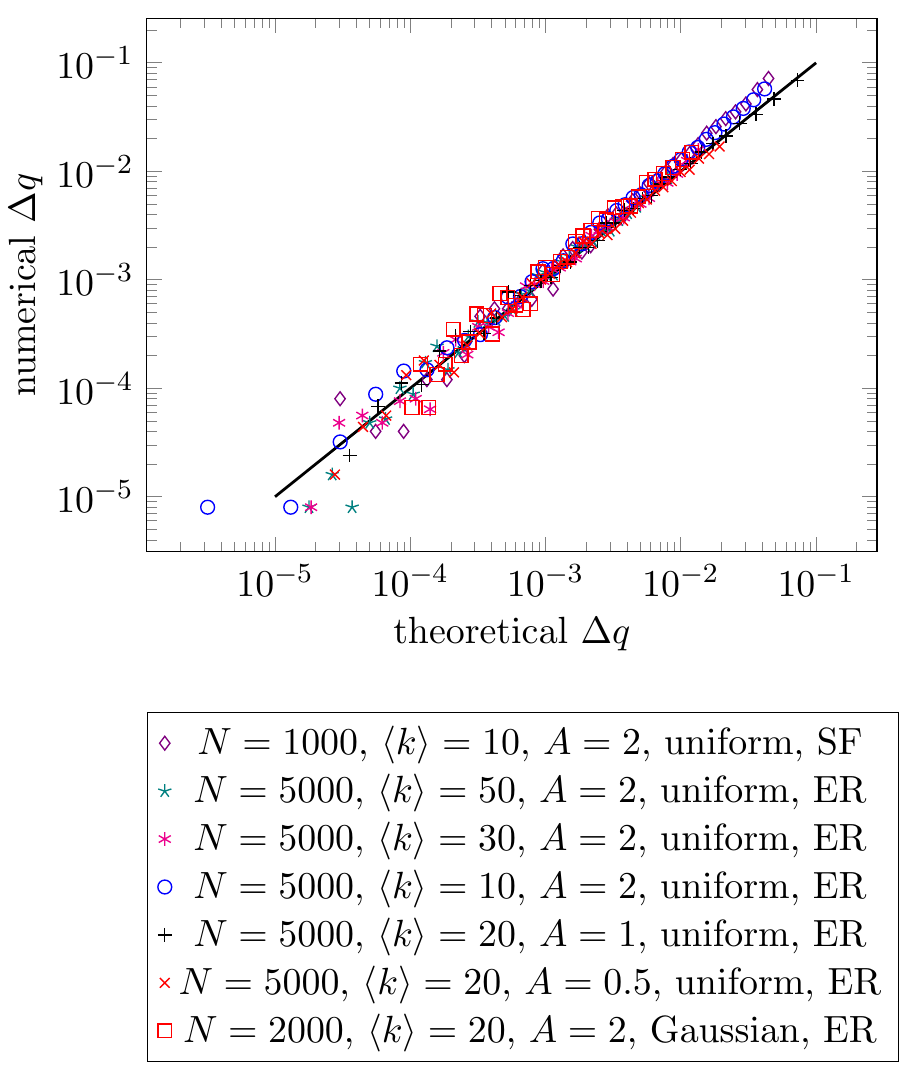} 

\caption{A comparison of the numerically determined width of the equilibria spectrum to Eq. (\ref{1overk}).   We see that the theory is very accurate even near a phase transition (these are the points with $\Delta q$ larger).   We have only shown a subset of the data for clarity.  In the data shown, scale free graphs have $\nu=2.5$ and Gaussian $P_v$ distributions have $\sigma=0.2$.}
\label{sweeps}
\end{figure}

As discussed earlier, we ran simulations of the binary decision model by first increasing $p$ from a state of all 1, and then decreased $p$ from states with many 0s.   This means that we can observe, quantitatively, the range of the spectrum of equilibria by observing the difference between the value of $q$ on the upward sweep versus the downward sweep.   Figure \ref{sweeps} shows that (\ref{1overk}) is \emph{quantitatively correct}, despite the incredibly simple theory we used.   

Indeed, it should not be so surprising that we found characteristics of glassy behavior, considering the similarity of this model to the random field Ising model.   Interestingly, it is unlikely this model is a true spin glass, due to its similarity to the random field Ising model where it was shown that the spin glass susceptibility is not divergent \cite{krzakala}.   This is perhaps more of a mathematical technicality than a physically meaningful statement:   in our model, so long as $\langle k\rangle$ is finite and $\alpha >0$, the binary decision model is \emph{always} a multistable ``glass".

\subsection{Multistability by the Cavity Method}
Since the multistability phenomenon is so fundamental to our model, let us predict multistability via a second approach:  roughly speaking, we will derive TAP-like equations for $x_v$ via the cavity method \cite{cirano}.   The cavity method is an extension of mean field theory which is used to understand disordered systems, and it essentially refers to picking a special node in the graph, called the ``cavity" node, and evaluating the probability that it is in each state, assuming that all of its neighbors feel the effect of the cavity's state exactly, with the remainder of the graph treated at mean field level.  Of course, this can be extended: an $m^{\mathrm{th}}$ order cavity method could take into account the reaction of the cavity on nodes up to $m$ edges away.  Note that the approach is also in the spirit of belief propagation on a tree \cite{mezard}.

To use the cavity method for our problem is straightforward: we need to make our mean field argument from earlier a bit more refined, so we will pick a special node $v$ (the ``cavity") and explicitly write \begin{equation}
x_v = \Theta\left(P_v-\frac{p}{h(q_v)}\right)
\end{equation}and explicitly determine $q_v$.   To find $q_v$, we need to determine the expected value of a state $x_u$, for each $u$ with $uv\in E$.     If we only use a first order cavity approximation, then $k_u-1$ of the edges of $u$ point to nodes with $x=q$, so we find \begin{equation}
x_u = \Theta\left(P_u - \frac{p}{h(q+(x_v-q)/k_u)}\right).
\end{equation}Averaging over the internal disorder \emph{and} graph structure it is straightforward to show that, assuming that $k$ is large, \begin{equation}
\mathrm{P}(x_u=1) = q_v = F\left(\frac{p}{h(q)}\right) + \frac{\alpha}{\langle k\rangle} (x_v-q) = q+ \frac{\alpha}{\langle k\rangle} (x_v-q).
\end{equation}Then we find the equation for the cavity node $v$, assuming again that $k$ is large enough that we can Taylor expand $h(q)$: \begin{equation}
x_v \approx \Theta\left(P_v - \frac{p}{h(q)} + \frac{ph(q)}{h^\prime(q)^2}\frac{\alpha}{\langle k\rangle}(x_v-q)\right).
\end{equation}Now averaging over the internal disorder of $P_v$, we find that it may be possible that both $x_v=0$ and $x_v=1$ are solutions.   The calculation of how likely this is is very simple: \begin{align}
\mathrm{P}\left(\frac{p}{h(q)}\left(1+\frac{h^\prime(q)}{h(q)}\frac{\alpha q}{\langle k\rangle}\right) > P_v> \frac{p}{h(q)}\left(1+\frac{h^\prime(q)}{h(q)}\frac{\alpha (q-1)}{\langle k\rangle}\right)\right) &\approx f\left(\frac{p}{h(q)}\left(1+\frac{h^\prime(q)}{h(q)}\frac{\alpha q}{\langle k\rangle}\right)\right) \frac{\alpha}{\langle k\rangle} \frac{ph^\prime(q)}{h(q)^2} \notag \\
&= \frac{\alpha^2}{\langle k\rangle} + \mathrm{O}(\alpha^3)   \label{1overk2}
\end{align}where the $\mathrm{O}(\alpha^3)$ terms correspond to terms proportional to the derivative of $f$.  This is precisely what we found earlier, neglecting the back reaction onto the remainder of the graph (via avalanches), up to the new terms which have arisen via the cavity method. 

Let us now perform the cavity method to higher orders:  in particular, let us assume that the graph is tree-like (at least locally), and keep track of all the nodes up to a graph distance of $n$ away from the cavity.   The tree-like approximation is very convenient, as it allows us to assume that each node feels the effects of the cavity node $v$ through exactly one neighbor.   We will start by considering the case of $n=2$ -- the generalization to larger $n$ will be very straightforward.    The state of the cavity is given by \begin{equation}
x_v = \Theta\left(P_v - \dfrac{p}{\displaystyle h\left(\frac{1}{k_v}\sum x_i\right)}\right)
\end{equation}and $x_1,\ldots, x_{k_v}$ denote the states of $v$'s neighbors.     The state of one of the neighbors of $v$, e.g. $x_1$, is given by \begin{equation}
x_1 = \Theta\left(P_1 - \dfrac{p}{\displaystyle h\left(\frac{x_v}{k_1} + \left(1-\frac{1}{k_1}\right)q_1\right)}\right),
\end{equation}where $q_1$ corresponds to the fraction of nodes (other than $v$) which are neighbors of node 1, and are in state 1.   But note that if $x_1$ is unknown, we already computed the formula for $q_1$ previously: \begin{equation}
q_1 = q+\frac{\alpha}{\langle k\rangle} (x_1-q)
\end{equation}where as before, $q$ is the fraction of nodes far from the graph.   Now, we write, assuming the number of nodes is large, as usual: \begin{equation}
x_1 = \Theta\left(P_1 - \frac{p}{h(q)}\left(1-\frac{h^\prime(q)}{h(q)}\left[\frac{\alpha}{\langle k\rangle}\left(1-\frac{1}{k_1}\right)(x_1-q) + \frac{x_v}{k_1}\right]\right)\right),
\end{equation}and using the same argument as before by finding the largest and smallest possible corrections to the effective value of $p$, we conclude that the probability that there is an equilibrium state with $x_1=1$, as well as one with $x_1=0$, given $x_v$, is given by \begin{equation}
\mathrm{P}(x_1=0\text{ or 1}|x_v) = \alpha \cdot \frac{\alpha}{\langle k\rangle}\left(1-\frac{1}{k_1}\right)+ \mathrm{O}(\alpha^3)
\end{equation}Note that we have not yet allowed for fluctuations in $x_0$, which we have assumed is fixed.   Also, given this formula, it should be fairly clear that we could have guessed this answer a priori from what we found before, under the assumption that one node $x_v$ is fixed.   

Finally, let us return to the question of interest:  the probability that $x_v$ can be in both states.   We need to compute the largest and smallest possible values of $q_v$:\begin{subequations}\begin{align}
\mathrm{P}(x_1=1|x_v=1) &= q + \alpha \left[\frac{\alpha}{\langle k\rangle}\left(1-\frac{1}{k_1}\right)(1-q) + \frac{1-q}{k_1}\right] \\
\mathrm{P}(x_1=0|x_v=0) &= q + \alpha \left[\frac{\alpha}{\langle k\rangle}\left(1-\frac{1}{k_1}\right)(-q) - \frac{q}{k_1}\right]
\end{align}\end{subequations}Note that the formula above neglects some of the smaller corrections due to derivatives in $f$, e.g., although these could easily be carried through if needed.  We may finally average over the degree distribution, and replace $1/k_1$ with $1/\langle k\rangle$.   Using these equations to find upper and lower bounds on $q_v$, we finally can conclude, using the same logic as in our earlier computation, that \begin{equation}
\mathrm{P}(x_v = 0\text{ or 1}) = \alpha \left[\frac{\alpha}{\langle k\rangle} + \frac{\alpha^2}{\langle k\rangle}\left(1-\frac{1}{\langle k\rangle}\right)\right] \approx \frac{\alpha^2(1+\alpha)}{\langle k\rangle}.
\end{equation}Note that this is also the width, $\Delta q$, of equilibria, up to second order in the cavity method.

Now, let us extend this computation to higher orders.   If we have found the width of the spectrum accounting for all nodes up to a distance $n$ away, assuming the graph is a tree, we can easily determine the width of the spectrum accounting for all nodes a distance $n+1$ away from $x_v$, by simply treating the neighbors of $x_v$ as the cavities and using the result for $n$.   An analogous computation to the one we did above shows that: \begin{equation}
\Delta q^{(n+1)} = \alpha\left[\frac{\alpha}{\langle k\rangle} + \Delta q^{(n)}\right].
\end{equation}Using the results for $\Delta q^{(1)}$ and $\Delta q^{(2)}$ from before, it is clear that \begin{equation}
\Delta q^{(n)} = \frac{1}{\langle k\rangle} \sum_{k=1}^n \alpha^{1+k}.
\end{equation}Summing this series to infinite order, we find that \begin{equation}
\Delta q^{(\infty)} = \frac{\alpha^2}{(1-\alpha)\langle k\rangle},
\end{equation}assuming that $\alpha<1$ so that the series is summable.    Thus, we see that the cavity method recovers the result we found using simple logic earlier.   

\subsection{Robustness to Fluctuations}
An important consequence of the spectrum of equilibria is that the macroscopic model is \emph{robust} against small perturbations in the external field $p$.   To estimate the size $\epsilon$ of a perturbation in the external field $p$ required to change the macroscopic state (the value of $q$),\footnote{It is possible that the equilibrium that we are at is at an outer edge of the spectrum, where this argument does not work, but for most equilibria (which are in the center of the spectrum) it works fine.} we use the fact that \begin{equation}
\frac{\mathrm{d}q(p)}{\mathrm{d}p} \sim \frac{1}{\epsilon} \frac{\alpha^2}{(1-\alpha)\langle k\rangle}.  \label{slope}
\end{equation}(\ref{slope}) follows directly from basic graphical considerations, by considering (\ref{qeq}) and using that if the slope of the mean field curve is known, then the fluctuations in both the $q$ and $p$ directions in $F(p/h(q))$ must be related.    Taking an implicit derivative of (\ref{qeq}), we find that \begin{equation}
\frac{\mathrm{d}q(p)}{\mathrm{d}p} = -\frac{\alpha}{1-\alpha} \frac{h(q)}{ph^\prime(q)},
\end{equation} which implies that \begin{equation}
\epsilon \sim \frac{\alpha}{\langle k\rangle}  \frac{ph^\prime(q)}{h(q)}.
\end{equation}For a generic model, the $h$-dependent factor in $\epsilon$ is likely O(1), and so the dominant feature is the $\alpha$ and $\langle k\rangle$ dependence.   We see that the model becomes more robust against small price fluctuations as we approach a critical point (where $\alpha=1$), although this is a linear approximation, as we assumed that the fluctuations were small enough that we could neglect terms beyond the first derivative in $F(p/h(q))$, in (\ref{slope}).   Thus, (\ref{slope}) will break down as nonlinear effects become important as we approach the critical point.

\subsection{Suppression of Discontinuous Phase Transitions}
What happens to the fluctuations of multistability as we approach the critical point, which mean field theory will in general take as a discontinuous phase transition?   Here, the simple linear arguments we used above begin to fail.   The $1/\langle k\rangle$ fluctuations can become so large on graphs with ``small" values of $\langle k\rangle$ that they can in fact remove the discontinuity in the phase transition.

  We can use a simple mean field theoretic argument to justify this.    Let $q_k$ denote the fraction of nodes with $k$ edges in state 1, and $r$ the probability that an edge points to a node in state 1:  \begin{equation}
r \equiv \frac{1}{\langle k\rangle} \sum_k k\rho_k q_k.  \label{req}
\end{equation}  A more realistic expression for $q_k$ is given by \begin{equation}
q_k = \sum_{m=0}^k  \frac{k!}{m!(k-m)!} r^m (1-r)^m  F\left(\dfrac{p}{\displaystyle h\left(\frac{m}{k}\right)}\right).   \label{realqk}
\end{equation}For simplicity, let us suppose that $k$ is large enough that binomial distribution is well approximated by a Gaussian distribution, so we have \begin{equation}
q_k \approx \int\limits_{0}^1 \frac{\mathrm{d}q}{\sqrt{2\pi k^{-1} r(1-r)}} \mathrm{e}^{-k(q-r)^2/2r(1-r)} F\left(\frac{p}{h(q)}\right).
\end{equation}
Certainly when $k$ gets very large, the Gaussian collapses to a $\delta$ function and we obtain $q_k=F(p/h(r))$, independently of $k$.   However, suppose that the length scale $q_F$ of fluctuations in $F(p/h(q))$ is comparable to $\sqrt{k^{-1}r(1-r)}$.   Then we see above that fluctuations in the number of nodes which are actually in state 1 may smooth out fluctuations in $F$, destroying the phase transition!

Since (\ref{realqk}) is far too complicated to analyze in the regime of interest, where $k$ becomes small, we resort to estimating numerically whether or not there will be a phase transition.   To do this we begin by determining self-consistent values of $r$, as calculated by using (\ref{req}) and (\ref{realqk}).    It is simplest to imagine doing this by plotting the function $r$, and then the function determined by (\ref{req}) and  (\ref{realqk}), and looking for intersections.   The reason this is the preferred method is because we have not yet taken into account the $1/\langle k\rangle$ fluctuations.  A crude way of doing this is simply adding $1/2\langle k\rangle$ to the right hand side of (\ref{req}), as we can consider the multistability as being caused by fluctuations in the value of $F(p/h(q))$ of  $\pm \alpha/2\langle k\rangle$.    As the previous sections have suggested, this effect can become quite serious near a phase transition.  Coupled with the effect of few edges smoothing out the weighting function $F(p/h(q))$ in the mean field equation, we find that theoretically we can see the disappearance of discontinuities in the phase transitions.    

\begin{figure}[here]
\centering

   \includegraphics{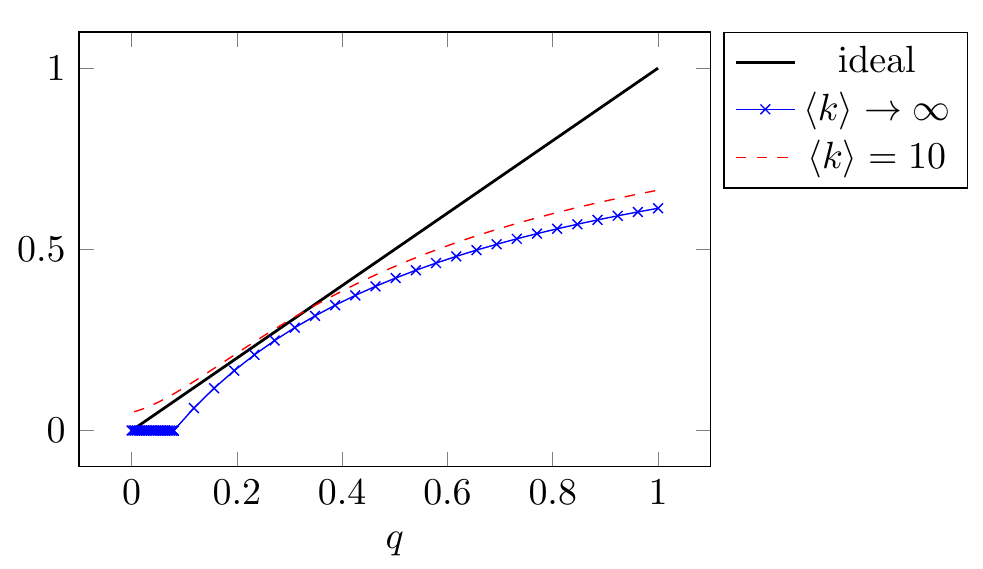}

\caption{A sample of how small network effects alter the mean field equation, assuming that $F(P)$ corresponds to the uniform distribution, the network is Erd\"os-R\'enyi, and $h(q)=1+2q$.   The given value of $p=1.16$ is quite close to the mean field critical point $p_{\mathrm{c}}=1.125$.   We have added the $1/2\langle k\rangle$ correction to the finite $\langle k\rangle$ curve.  Note that while for the ideal case, the only solution is $q=0$, the small $\langle k\rangle$ curve still intersects the line at a positive value of $q$ -- i.e., the phase transition has not occurred yet.}
\label{fig8}
\end{figure}

This gives us a crude way to estimate numerically whether a graph will admit a discontinuous or continuous phase transition.  In Table \ref{thetab} we perform the procedure above, assuming that $F(P)$ is the uniform distribution, and estimate the value of $\langle k\rangle$ for Erd\"os-R\'enyi or scale free graphs for which we should see the onset of shocks and discontinuous transitions.   A sample of what the adjusted mean field curve looks like, near the mean field critical point, with low $\langle k\rangle$  is also shown in Figure \ref{fig8}.   Numerical estimates suggest this transition should be fairly sharp and we observed this numerically as well.   Because we do not have a precise way of determining whether a phase transition has been discontinuous, other than simply to observe the size of the change in $q$ at each step forward in $p$, these results should be taken as at most semi-quantitative.  However, they do suggest that we have correctly identified the mechanism for the disappearance of discontinuous phase transitions.
\begin{table}[here]

\centering
\begin{tabular}{|c|c|c|c|}\hline
graph type &\ $A=2$ &\ $A=3$ &\ $A=4$ \\ \hline 
scale free, $\gamma=2.5$ &\ 20/\color{blue} 20 &\ 14/\color{blue} 14 &\ 10/\color{blue} 10 \\ 
scale free, $\gamma=3$ &\ 18/\color{blue} 20 &\ 12/\color{blue} 14 &\ 10/\color{blue} 10 \\ 
scale free, $\gamma=4$ &\ 18/\color{blue} 20 &\ 12/\color{blue} 14 &\ 10/\color{blue} 10 \\
Erd\"os-R\'enyi &\  18/\color{blue}18 &\ 13/\color{blue} 13 &\ 10/\color{blue} 10 \\ \hline
\end{tabular}

\caption{The theoretical (blue) vs. numerical (black) values for $\langle k\rangle$ at which we expect to see the onset of discontinuous phase transitions (into the $q=0$ state) for the uniform distribution.  Because these predictions require a precise understanding of multistability at the critical point, they should not be taken too seriously:  we believe a standard deviation of $\pm 2$ on each data point is not unreasonable (as we can only make scale free graphs with even $\langle k\rangle$ with our fast algorithm).}
\label{thetab}
\end{table}

Of course, there remains the question of whether or not these fluctuations actually merge the two phases together.   This is a question beyond the scope of the simple arguments of this paper.  We suggest that the answer may be that it depends on the network/disorder/interactions.   For example, in the case with uniform $F(P)$, there is a very sharp discontinuous phase transition as the value of $p$ is lowered from above to below the maximum allowed $P_v$.   This transition is hard to remove because the coefficient of the $1/\langle k\rangle$ fluctuations is in fact 0 (as all nodes are in $x_v=0$), and so network fluctuations will not affect this transition.  Thus at least for these systems, we conclude that there must be 2 distinct phases, although one of the transitions between the two phases may be continuous.

\section{Conclusion}
We have presented the binary decision model and explored its major aspects:  understanding from both a macroscopic and microscopic view the mean field solutions and fluctuations of basic quantities, and then understanding the ``glassy" multistability phenomena and the consequences of a  spectrum of equilibria.   The simple form of the decision making, as well as the interactions, allowed us to nearly exactly solve this model with very simple, physically motivated arguments.   However, there are many questions about this model which are still open.   Firstly, it is unknown how robust the basic features of this model are to modifications:  for example, heterogeneity among nodes in the interactions $h(q)$, or ``thermal" random behavior among the nodes.  Secondly, while it is unlikely that the multistability effect is a true spin glass effect, it is an interesting question whether or not this model's dynamics at ``finite temperature" would exhibit aging, another characteristic feature of a glass.  Thirdly, an investigation of the model on non-random graphs such as hypercubic lattices, or on graphs with many loops, may help to shed some light on the nature of multistability, as we mentioned earlier.

It  appears that on heterogeneous networks, the binary decision model has \emph{fundamentally different} behavior from the random field Ising model, despite the similarity in the motivation between the two models.   One can intuitively see this as follows:  for the random field Ising model with coupling $J$, using the additive formulation, the equation of state is \begin{equation}
x_v = \Theta(P_v + Jk_vq_v - p).
\end{equation}This equation explicitly depends on $k_v$, the number of edges of $v$.   Therefore, we expect the mean field equations to depend on graph structure, unlike in the binary decision model.   Further study of qualitative differences in behavior between these two models, as well as modified fiber bundle models with the ability to ``regenerate" nodes as $p$ is decreased, is a worthwhile direction for further study.

Finally, it is important to understand the ways in which this model can be tested against empirical data.   In general, this is quite challenging, but let us conclude by discussing some features of this model which may be observable.    One of the most interesting features of our model is the remarkable robustness of our results against the degree distribution of the graph.   For example, the probability that a node is involved in an avalanche is independent of its degree.   Secondly, given $\langle k\rangle$, one could compare the width of the spectrum of equilibria to the typical size of avalanches, and should find the same value of $\alpha$ from both measurements.  There are two important drawbacks to this approach.  Firstly, trying to show the robustness of opinion dynamics against graph structure may require better knowledge of the social graph than can be obtained.   Secondly, it is unclear how robust our formulas are to model modifications:  for example, correlations between $P_v$ and $k_v$ may restore the degree distribution into many of our results.

The most important signature to look for in empirical data is ``multistability."    This is unlikely to be a peculiarity of our model's precise formulation, and discovery of such behavior would confirm the importance of emphasizing glassy features in opinion models.     We leave further understanding of experimental signatures of this model, and related ones, to future work.

\section*{Acknowledgements}\addcontentsline{toc}{section}{Acknowledgements}
We would like to thank Jean-Phillipe Bouchaud and Ariel Amir for helpful comments, as well as the suggestions of the anonymous referees.

A.L.  is supported by the Purcell Fellowship at Harvard. CH.L.   is supported by the Agency of Science, Technology and Research of Singapore.

\bibliographystyle{plain}
\addcontentsline{toc}{section}{References}
\bibliography{marketbib}

\end{document}